# Stacking-Engineered Switchable Altermagnetism in Topological FeSe bilayer systems

Jie Li,[1,*] Shifang Li,[1] Mengyang Zhang,[1] Pan Zhou,[2,*] Jianxin Zhong,[1,*] and Ruqian Wu[3,*]

[1]*Institute for Quantum Science and Technology, Shanghai University, Shanghai 200444, China*

[2]*Hunan Provincial Key laboratory of Thin Film Materials and Devices, School of Materials Science and Engineering, Xiangtan University, Xiangtan 411105, China*

[3]*Department of Physics and Astronomy, University of California, Irvine, CA 92697- 4575, USA*

Altermagnetism and topological insulators represent two of the most transformative frontiers in modern condensed matter physics, spintronics, and quantum information science. Bringing these two paradigms together opens a largely unexplored route toward fundamentally new quantum phenomena. Here, we predict a topological altermagnetic phase in bilayer tetragonal Fe-based superconductors and reveal it as a highly tunable platform for valley-polarized anomalous Hall physics. Based on first-principles calculations, we show that the characteristic spin-splitting and valley polarization can be effectively tuned via applied strain. Moreover, the resulting valley-polarized anomalous Hall conductivity can be manipulated by shifting the Fermi level. These findings reveal a powerful route for controlling altermagnetism in topological materials and identify a realistic material platform for its experimental realization and technological exploitation.

*Corresponding authors: lij@shu.edu.cn; zhoupan71234@xtu.edu.cn; jxzhong@shu.edu.cn and wur@uci.edu

## I. INTRODUCTION

Altermagnets (AMs), recently established as a third class of magnetic materials, uniquely combine the advantages of both ferromagnetic and antiferromagnetic ordering and have rapidly emerged as a central focus in spintronics and quantum materials research [1-5]. In conventional collinear antiferromagnets, spin degeneracy of electronic bands is protected by the combinations of spatial symmetries such as inversion ($P$) or translation ($t$) and time-reversal symmetry ($T$), rendering spin splitting accessible primarily through strong spin-orbit coupling (SOC) or external Zeeman fields. By contrast, altermagnets feature momentum dependent spin splitting and maintain zero net magnetization due to the absence of $PT$ or $tT$ symmetry. Thus, altermagnets exhibit time-reversal symmetry-breaking magneto responses, vanishing stray fields and high-frequency spin dynamics [6], making them highly attractive for next generation information storage and processing technologies [7-9]. Although the spin splitting in altermagnets is fundamentally independent of SOC, most realistic materials possess appreciable SOC, particularly in transition metal-based systems. Because strong SOC underpins a wide range of topological phase, its interplay with altermagnetic order is poised to generate fundamentally new quantum phenomena [10-11].

To date, several strategies have been proposed to engineer altermagnetism from conventional antiferromagnets by deliberately breaking $P$-symmetry or $t$-symmetry, including interlayer magnetoelectric coupling [8,12,13], external electric fields [14-16], or crystal structure design [17-19]. Despite this progress, achieving precise and reversible control over altermagnetic properties, a prerequisite for practical device application, remains a long-standing challenge, and the switching behavior of altermagnets has rarely been investigated [16,20]. Thus, the quest of novel topological altermagnets together with efficient and versatile tuning mechanisms has emerged as a pressing and multidisciplinary research frontier for the development of controllable altermagnetic devices.

In this work, we explore a general route for engineering topological altermagnets by stacking bilayer antiferromagnetic topological materials. Using a tight-binding model, we reveal that the spin-space symmetry breaking can induce altermagnetism even in stacking parity-symmetric bilayer antiferromagnetic structures. Guided by this principle, we propose a realistic vdW bilayer platform based on tetragonal Fe-based superconductors (FeSe and FeTe) as prototypical model system (BL-FeSe or BL-FeTe). Density functional theory (DFT) calculations indeed confirmed that the effectiveness of modulating spin-space symmetry to switch altermagnetism in BL-FeSe and BL-FeTe. Moreover, we demonstrate that the resulting spin-splitting and valley polarization can be efficiently tuned through the application of out-plane and in-plane strain. These findings establish a practical and versatile strategy for realizing and controlling topological altermagnetism, with direct implication for next generation spintronic devices.

## II. METHODOLOGY

In this work, all the density functional theory (DFT) calculations are carried out with the Vienna ab-initio simulation package (VASP) at the level of the spin-polarized generalized-gradient approximation (GGA) with the functional developed by Perdew-Burke-Ernzerhof [21]. For the interaction between valence electrons and ionic cores, the framework of the projector augmented wave (PAW) method is adopted [22,23]. The Hubbard U of 2.0 eV is set to the electron correlation in the d-shells of Fe cores [24]. To well describe the dispersion forces, the vdW correction (DFT-D3) is included [25]. In DFT calculations, the energy cutoff for the plane wave basis expansion is set to 500eV, and the criterion for total energy convergence is set to $10^{-5}$ eV. All atoms are fully relaxed using the conjugated gradient method for the energy minimization until the force on each atom becomes smaller than 0.01 eV/Å.

## III. RESULTS AND DISCUSSION

**The tight-binding model prediction**. The similarity in magnetic order, namely, the presence of opposing spin sublattices, makes it possible to design altermagnets starting from conventional antiferromagnets. To illustrate this idea, we consider a monolayer collinear antiferromagnet with $D_{4h}$ symmetry (see in Fig. 1a) as an example, and construct a tight-binding (TB) model:

$$H = \sum_{i,j} t_{ij} c_i^\dagger c_j + \varepsilon \sum_i c_i^\dagger c_i + j \sum_i c_i^\dagger \tau_z \sigma_z c_i + h.c. \quad (1)$$

where $c_i^\dagger$ and $c_i$ are the creation and annihilation operators at site *i,* respectively, and τ and σ denote Pauli matrices acting on the sublattice and spin degrees of freedom. $t_{ij}$ represents the hopping between site *i* and *j*, and only the nearest-neighbor hopping ($t_1$) and next-nearest-neighbor hopping ($t_2$) are considered in monolayer collinear antiferromagnet. $\varepsilon$ and $j$ characterizes the on-site energy and AFM exchange interaction, respectively (see the supporting materials for details [26]). Consistent with expectations, TB model simulations demonstrate that the band structure exhibits spin-degeneracy throughout the Brillouin zone, protected by PT symmetry (see in the left of Fig. 1d). Nevertheless, when an external potential of the form:

$$H_{ext} = \Delta \sum_i c_i^\dagger c_i \quad (2)$$

is added to Eq. 1, an additional anisotropic lattice potential is introduced. As a result, the opposite spin sublattices experience opposing external potential fields, enabling the desired spin splitting, as illustrated on the right side of Fig. 1d. Consequently, achieving a transition from an antiferromagnet to an altermagnet hinges on introducing inequivalent external potentials that break the relevant

spin–space symmetries. In this work, we realize this goal through stacking engineering, starting from a vdW homostructure composed of two antiferromagnetic layers. First, we consider a simplified scenario in which only magnetic atoms are included (labeled M1, see in Fig. 1b). The corresponding bilayer Hamiltonian can be written as:

$$H = \left(\sum_{i,j} t_{ij} c_i^\dagger c_j + \sum_i \varepsilon_i c_i^\dagger c_i + J \sum_i c_i^\dagger \tau_z \sigma_z c_i\right)_{L1} + \left(\sum_{i`,j`} t_{i`j`} c_{i`}^\dagger c_{j`} + \sum_{i`} \varepsilon_{i`} c_{i`}^\dagger c_{i`} + J \sum_{i`} c_{i`}^\dagger \tau_z \sigma_z c_{i`}\right)_{L2} + \left(\sum_{\langle i,L1\rangle\langle i`,L2\rangle} t_{ii`} c_i^\dagger c_{i`}\right)_{inter-L} + h.c. \quad (3)$$

where the first, second and third terms represent the contributions of the 1st layer (L1), 2nd layer (L2) and interlayer coupling (inter-L), respectively (see the supporting materials for details [26]). TB model simulations demonstrate that, in this case, breaking spin-space symmetry alone is insufficient to induce altermagnetism (see in Fig. 1e). This indicates the crucial role of the crystal environment provided by nonmagnetic atoms. Motivated by this insight, we construct two bilayer models that explicitly incorporate nonmagnetic atoms, based on monolayer antiferromagnets that preserve either P-symmetry or mirror-symmetry (labeled M2 and M3, respectively; see in Fig. 1c and Fig. S4a [26]). Upon including the hoppings between magnetic and nonmagnetic sites, our TB analysis reveals that spin splitting occurs exclusively in the M2 structure due to spin-space symmetry breaking (M2-S1). Furthermore, by reversing the spin order in one layer (M2-S2), the spin-space symmetry can be selectively switched on or off, thereby enabling or disabling the $PT$ symmetry that enforces degeneracy between opposite-spin bands (their symmetry analysis as shown in theTable I). Consequently, this M2-based bilayer homostructure provides a versatile and tunable platform for altermagnetism, opening new avenues for reconfigurable spintronic devices. (see the supporting materials for more detail of TB model simulations [26]).

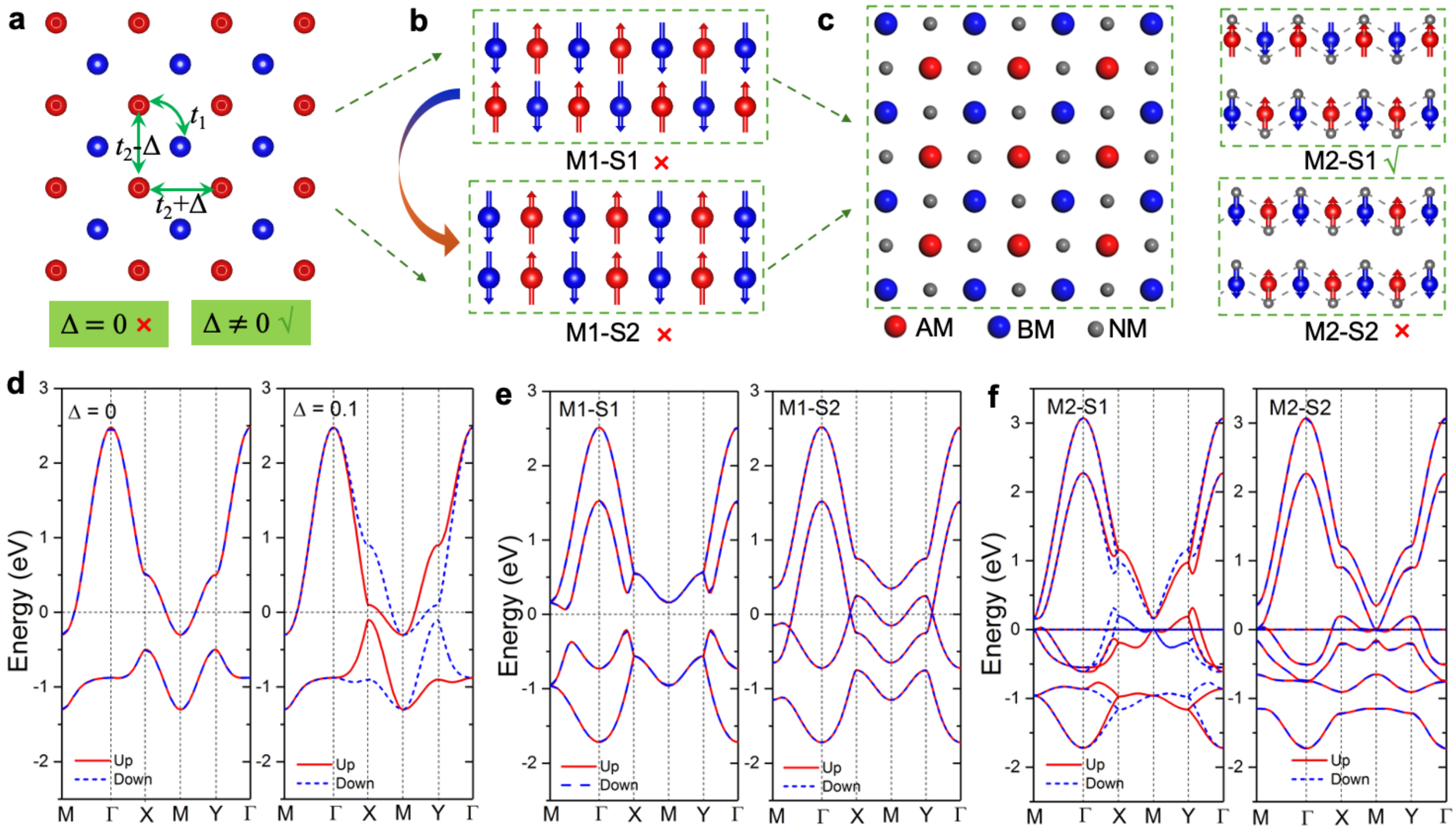

FIG 1. (a) Schematics of a monolayer collinear antiferromagnets with $D_{4h}$ symmetry. (b) The side view of stacking bilayer antiferromagnets with only magnetic atoms included under two different interlayer coupling states (M1-S1 and M1-S2). (c) (left) Top view of bilayer antiferromagnets with magnetic atoms in two sites (AM, and BM) and nonmagnetic atoms (NM) included (M2), (right) the corresponding side view with two different interlayer coupling states (M2-S1 and M2-S2). (d-f) Their corresponding band structures from TB model simulations.

Table I. The magnetic space group (MSG), spin space group point groups (SSG) and typical symmetry operation of M1, M2 and M3, respectively.

| | M1-S1 | M1-S2 | M2-S1 | M2-S2 | M3-S1 | M3-S2 |
|---|---|---|---|---|---|---|
| MSG | $P_C4/nbm$ | $P_C4/mbm$ | $P4'/nm'm$ | $P4'/n'm'm$ | $P_C4/nbm$ | $P_C4/mbm$ |
| SSG | $P^{1}4/^{1}n^{1}m^{1}m^{-}$ $^{1}(1/2\ 1/2\ 0)^{\infty m}1$ | $P^{1}4/^{1}m^{1}m^{1}m^{-}$ $^{1}(1/2\ 1/2\ 0)^{\infty m}1$ | $P^{-1}4/^{1}n^{1}m^{-}$ $^{1}m^{\infty m}1$ | $P^{-1}4/^{-1}n^{1}m^{-}$ $^{1}m^{\infty m}1$ | $P^{1}4/^{1}n^{1}m^{1}m^{-}$ $^{1}(1/2\ 1/2\ 0)^{\infty m}1$ | $P^{1}4/^{1}m^{1}m^{1}m^{-}$ $^{1}(1/2\ 1/2\ 0)^{\infty m}1$ |
| Symmetry | PT & tT | PT & tT | $C_4$T | PT | PT & tT | PT & tT |

**DFT verifications**. Guided by TB analysis presented above, an ideal building block for realizing altermagnetism is a monolayer AFM material with *P*-symmetry. Through a systematic survey of candidate topological AFM materials, we identify tetragonal iron-based superconductors (FeSe and FeTe) as particularly promising platforms, as they naturally host both *P*-symmetry and AFM magnetic order. Notably, single atomic layer of FeSe has already been successfully grown epitaxially on $SrTiO_3$ substrates [27,28], suggesting that there are no fundamental technological barriers for synthesizing bilayer systems based on FeSe or FeTe. Density functional theory (DFT) calculations yield in-plane lattice constants of 3.72Å for FeSe and 3.78Å for FeTe, which are slightly smaller than those of their bulk counterparts. In this work, we therefore focus on homojunction systems, particularly bilayer FeSe or FeTe, owing to their superior experimental feasibility. Structural optimizations with the vdW correction included in DFT functional shows that both FeSe and FeTe prefer an A-A stacking configuration, and the corresponding interlayer distances are 2.59 Å for FeSe and 2.82Å for FeTe, respectively. To further assess their dynamic stability, we perform ab initio molecular dynamics (AIMD) simulations. The results show that no noticeable structural deformation for a 5×5 supercell containing 200 atoms after 5000 MD steps (10ps) at 300K. Moreover, the total energies fluctuate smoothly around the equilibrium value without abrupt changes (see Fig. S5 [26]), confirming that bilayer FeSe or FeTe are dynamically and thermally stable at least up to room temperature.

As shown in Figs. 2a and 2b, A-A stacked antiferromagnetic FeSe (FeTe) has the *P*- and $C_4$-symmetries in real space. Upon incorporating spin group, there are two interlayer magnetic orderings (BL-FeSe-S1 and BL-FeSe-S2). In BL-FeSe-S1, the *P*-symmetry only keeps in the same spins rather than opposite spins, i.e., its *PT*-symmetry is broken, whereas in BL-FeSe-S2 the *PT* symmetry is preserved by inversion connecting opposite-spin states. To verify whether these bilayer models indeed exhibit the predicted switchable altermagnetic behavior, we first examine their band structure through DFT calculations. As shown in Figs. 2c and 2d, pronounced momentum dependent spin-splitting is clearly observed in BL-FeSe-S1, while it is absent in BL-FeSe-S2. Notably, the altermagnetic state in bilayer FeSe can be reversibly switched by simply flipping the spin order in one FeSe layer. Furthermore, DFT calculations indicate that BL-FeSe-S1 is energetically favored by 2.4meV/supercell, implying that the switching between these two magnetic configurations can be readily achieved via external stimuli, such as by strain [29,30], external electric fields [31,32], or proximity effect [33]. Analysis of the projected band structure (see in Fig.

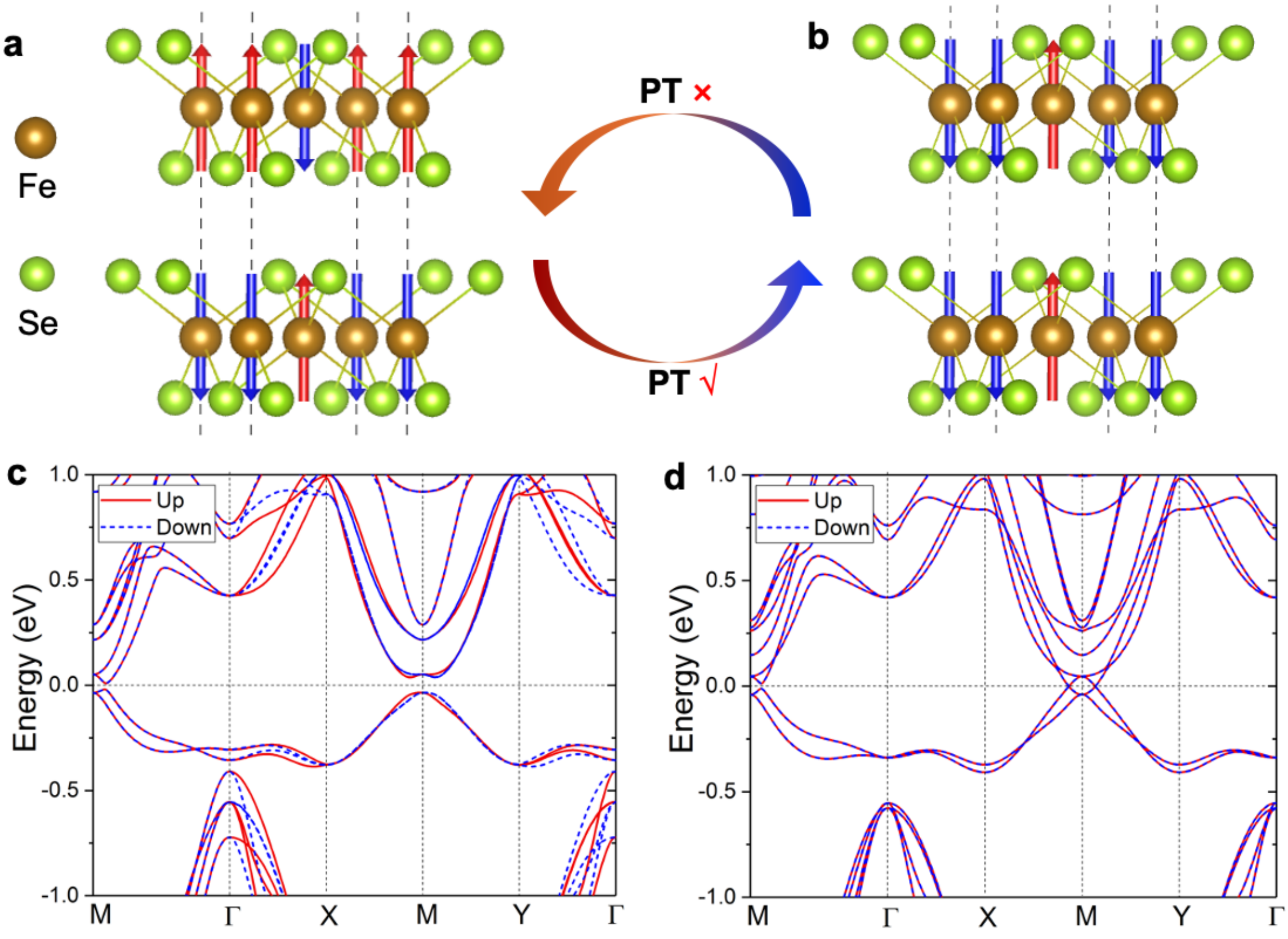

FIG. 2. (a)-(b) Schematics of bilayer FeSe with two different interlayer couplings (BL-FeSe-S1 and BL-FeSe-S2). (c)-(d) Their corresponding band structures by DFT calculations.

S6[26]) reveals that the bands from two layers FeSe are highly layer-degenerate due to the weak interlayer interactions. Consequently, the observed spin-splitting between spin-up and spin-down states originates primarily from an individual FeSe layer, while interlayer interaction offers the inequivalent external potentials required to stabilize the altermagnetic phase.

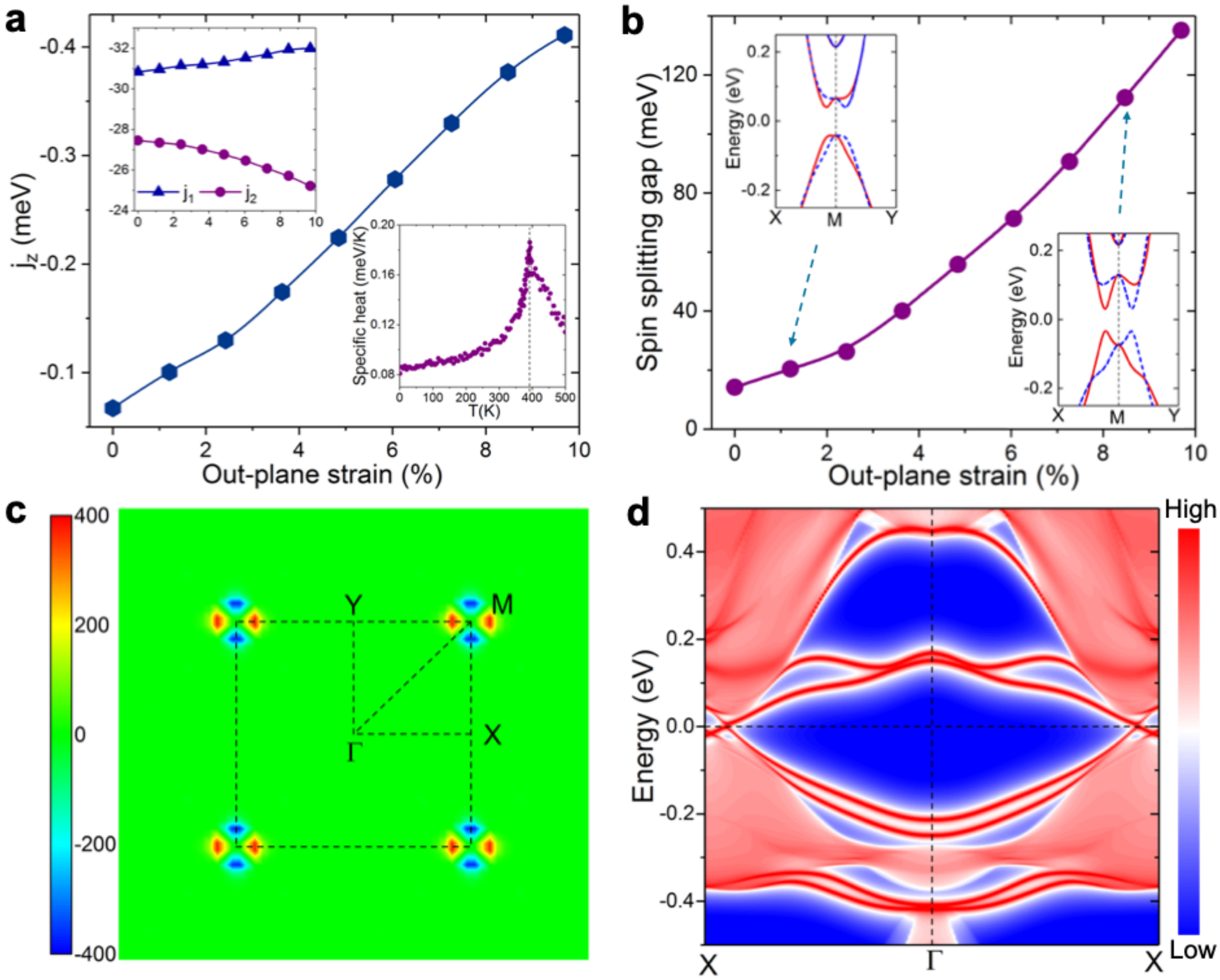

FIG. 3. (a)-(b) The interlayer and intralayer (in the upper-left panel of a) exchange coupling and spin splitting gap of bilayer FeSe as a function of out-plane strain (The specific heat of bilayer FeSe as function of temperature T as shown in the right inset of (a)). (c)-(d) The distribution of Berry curvature in the 2D Brillouin zone and the corresponding 1D band structure for a BL-FeSe-S1 ribbon, respectively.

To further assess the stability of the altermagnetic ground state, we calculated the corresponding Néel temperature using Monte Carlo simulations based on the classical Heisenberg model:

$$H = H_0 - J_1 \sum_{<i,j>_{intra}} S_i \cdot S_j - J_2 \sum_{<i,j>_{intra}} S_i \cdot S_j - J_z \sum_{<i,j>_{inter}} S_i \cdot S_j \ , \tag{4}$$

Here, $J_1$ and $J_2$ represent the nearest, next-nearest neighbor intralayer exchange interaction parameters, respectively, while $J_z$ denotes the nearest neighbor interlayer exchange interaction. The spin quantum number for Fe atoms, $S_i$, is taken as 3/2, and all exchange parameters are evaluated within a $3 \times 3$ supercell. Based on DFT calculations for four magnetic states as shown in Fig. S7 [26], the fitting exchange interaction parameters are extracted by fitting the total energies. As shown in Fig. 3a, the intralayer exchange interactions are found to exceeds the interlayer coupling by nearly two orders of magnitude, highlighting the quasi-two-dimensional magnetic nature of bilayer FeSe. Importantly, BL-FeSe has a Néel temperature up to 392K due to the huge intralayer exchange interaction parameters (-30.83meV, -27.45meV for $J_1$ and $J_2$, respectively), in consistent with previous works [34].

Notably, the spin splitting in BL-FeSe originates from the interlayer exchange interaction, suggesting that the altermagnetic characteristics can be effectively strengthened by tuning $J_z$. To demonstrate the viability of this approach, we apply out-plane strain. The DFT calculations reveal that the interlayer exchange interaction exhibits a much stronger response to out-plane strain than the intralayer exchange interactions as shown in Fig. 3a. Specially, $J_z$ increases nearly linearly with increasing out-plane strain, showing an enhancement of up to 510% when the out-plane strain is raised by 10%. As a direct consequence, the spin-splitting gap in BL-FeSe-S1 is markedly enlarged, from 14.2 meV to 135.1 meV, as shown in Fig. 3b and Fig. S8 [26].

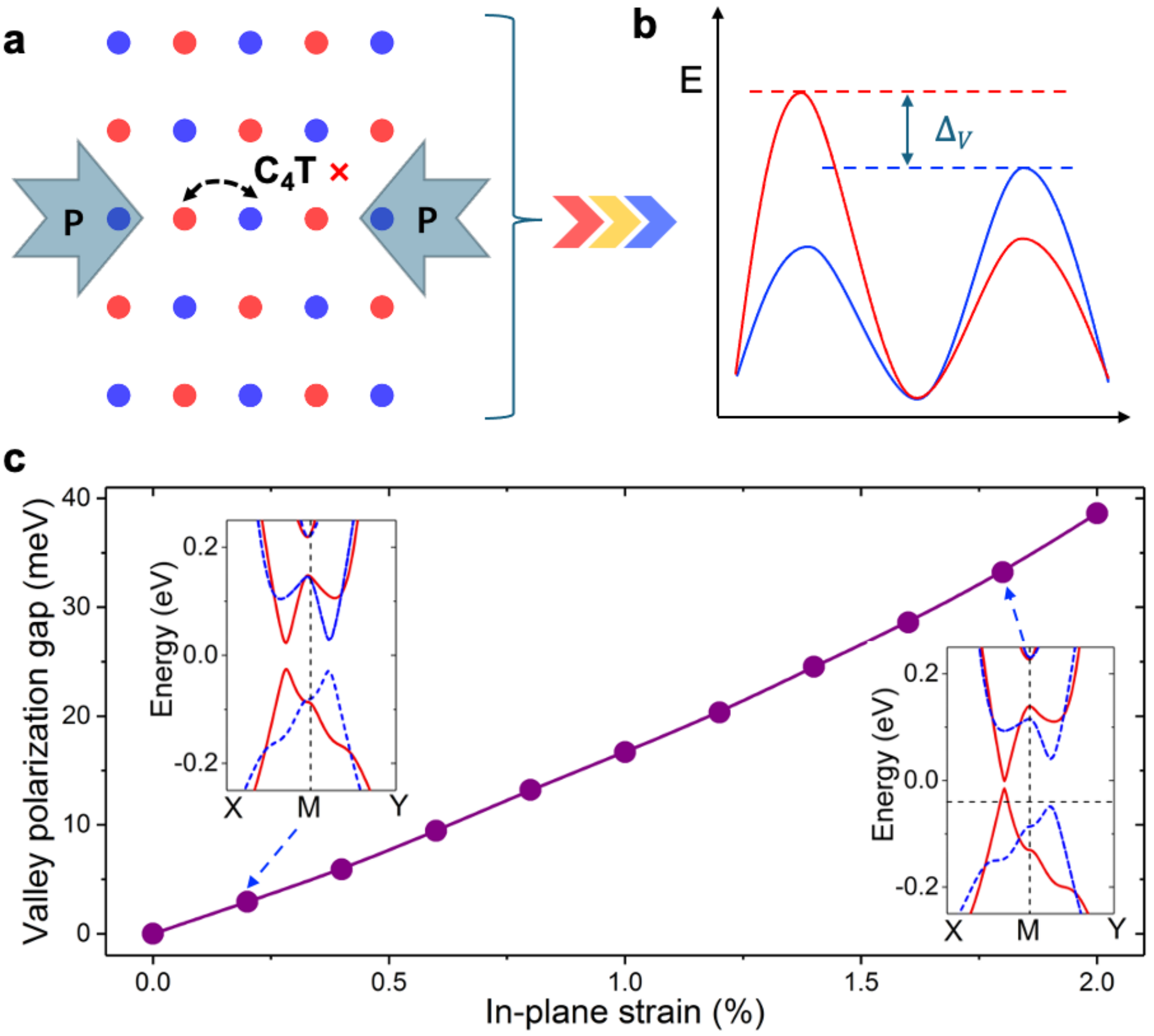

FIG. 4. (a)-(b) Schematic diagrams of generating valley polarization in altermagnets by breaking its $C_4T$ symmetry via a compressive in-plane uniaxial strain. (c) The valley splitting gap ($\Delta_V$)of BL-FeSe-S1 as a function of in-plane strain.

Because the coexistence of altermagnetism and nontrivial topology lies at the heart of quantum materials research, it is crucial to examine whether BL-FeSe-S1 retains topological characteristics of monolayer FeSe. To this end, we calculate its band structure including SOC (see in Fig. S9a [26]). Notably, SOC further enlarges the band gap near the M point from 72.6 meV to 89.5 meV, indicating enhanced topological robustness. Along the high-symmetry X-M-Y path in BL-FeSe-S1, the Berry curvature $\Omega(k)$ exhibits two pronounced peaks of opposite sign (see in Fig. S9b [26]), which are directly associated with altermagnetic spin valley polarization. Mapping the Berry curvature over the full Brillouin zone reveals four positive and four negative contributions localized near the zone corners in Fig. 3c. Consequently, the integration of Berry curvature over the Brillouin zone yields a vanishing Hall conductance, analogous to the quantum spin Hall effect and valley-polarized quantum Hall effect. To further confirm the intrinsic topological nature of BL-FeSe-S1, the $Z_2$ numbers is calculated. We may see that $Z_2$=1 by counting the positive and negative n-field numbers over half of the torus (see in Fig. S9c [26]). Furtherly, we construct a TB model using parameters extracted from the DFT band structure and compute the electronic spectrum of BL-FeSe-S1 nanoribbons. The emergence of robust edge states across the Fermi level in Fig. 3d provides direct and compelling evidence of its nontrivial topological character.

Clearly, lifting the degeneracy between the valleys near the $M$ point the along $\Gamma - X$ and $\Gamma - Y$ directions is the key requirement for generating a nonzero net Hall conductance, which is essential for device applications. Conventionally, valley degeneracy can be lifted by external magnetic field, magnetic doping or proximity coupling to magnetic materials. In BL-FeSe-S1, however, the valley degree of freedom and crystal symmetry are interlocked, the valley can be polarized by simply breaking the x-y crystal symmetry using in-plane uniaxial strain, as schematically illustrated in Fig. 4. DFT calculations confirm the effectiveness of this strategy. As shown in Fig. 4c and Fig. S10 [26], the valley polarization gap increase in an approximately linear fashion, reaching about 38 meV within a reasonable range of in-plane strain of less than $2\%$. The distribution of the Berry curvature in the Brillouin zone (see in Fig. S11 [26]) show a narrowed region along the Γ-X path under the compression. Once the valley degeneracy is lifted, a nonzero net Hall conductivity can be readily achieved by shifting the Fermi level through electrostatic gating or chemical doping. Specifically, when the Fermi level intercepts one valley, as indicated by the dashed line in the right inset of Fig. 4c, the positive Berry curvature contribution dominates, giving rise to a net Hall current, as illustrated in Fig. S11b.

## IV. CONCLUSION

In summary, we identify the AFM vdW BL-FeSe (BL-FeTe) homostructures as a compelling and experimentally viable platform for switchable topological altermagnetism, as established by complementary tight-binding modeling and first-principles calculations. The BL-FeSe homostructure simultaneously host robust nontrivial topology and altermagnetic order, with a Néel temperature exceeding room temperature. Remarkably, the altermagnetic spin-splitting can be dramatically amplified from 14.2 meV to 135.1 meV by increasing the interlayer exchange interaction via modest out-of-plane strain, revealing an efficient and purely structural control knob for altermagnetism. Furthermore, we demonstrate a symmetry driven route to valley polarization in altermagnets using in-plane uniaxial strain. Within experimentally accessible strain levels (<2%), the valley polarization gap increases nearly linearly, reaching 38 meV and enabling a valley-polarized anomalous Hall effect through simple Fermi-level tuning. Together, these results establish bilayer FeSe and FeTe as a rare example where altermagnetism, topology, strain engineering, and valley physics converge in a single material system. Our work not only introduces a powerful strategy for controlling altermagnetism in topological materials but also opens a realistic pathway toward field-free, reconfigurable spin- and valley-based devices, significantly expanding the landscape of magnetic topological quantum materials.

**Acknowledgements**

This work was partially supported by the National Natural Science Foundation of China (No. 12304089, 12374046), the Shanghai Science and Technology Innovation Action Plan (Grant No.

24LZ1400800), and the star-up foundation from Shanghai University. Calculations were partially performed computers at Shanghai Technical Service Center of Science and Engineering Computing, Shanghai University.

simulations of BL-FeSe and BL-FeTe, the projected band structure of BL-FeSe-S1, the four magnetic states, the band structure of BL-FeSe-S1 under out-plane strain, The Berry curvature **Ω**(k) distribution and n-field configuration of BL-FeSe-S1, the band structure of BL-FeSe-S1 under in-plane strain, the distribution of Berry curvature of BL-FeSe-S1 under in-plane strain of 2%.